\documentclass[twocolumn,showpacs,preprintnumbers,amsmath,amssymb]{revtex4}
\usepackage{graphicx}

\begin{document}

\title{Kondo-like behaviors in magnetic and thermal properties of single crystal Tm$_{5}$Si$_{2}$Ge$_{2}$ }

\author{J. H. Kim$^1$}
\author{S. J. Kim$^1$}
\author{C. I. Lee$^1$}

\author{M. A. Jung$^1$}
\author{H. J. Oh$^{1,2}$}
\author{Jong-Soo Rhyee$^3$}
\author{Younghun Jo$^4$}
\author{Hiroyuki Mitani$^{5,6}$}
\author{Hidetoshi Miyazaki$^{5}$}
\author{Shin-ichi Kimura$^{5,7}$}
\author{Y. S. Kwon$^1$}
\email{yskwon@skku.ac.kr}

\affiliation{$^{1}$ Department of Physics, Sungkyunkwan University,
Suwon 440-746, South Korea \\$^2$ Department of Ophthalmic Optics, Masan College, Masan 630-729, Republic of Korea \\$^3$ Inorganic Materials Group, Samsung
Advanced Institute of Technology, Yong-In 446-712, Republic of Korea
\\$^4$ Nano Materials Research Team, Korea Basic Science Institute, Daejeon 305-333, Korea \\$^5$ UVSOR Facility, Institute for Molecular Science, Okazaki 444-8585, Japan
\\$^6$ Graduate School of Engineering, Shinshu University, Nagano 380-8553, Japan
\\$^7$School of Physical Sciences, The Graduate University for Advanced Studies (SOKENDAI), Okazaki 444-8585, Japan}
\date{\today}

\begin{abstract}

We grew the single crystal of stoichiometric Tm$_{5}$Si$_{2.0\pm 0.1}$Ge$_{2.0\pm 0.1}$ using a Bridgeman method and performed XRD, EDS, magnetization, ac and dc magnetic susceptibilities, specific heat, electrical resistivity and XPS experiments. It crystallizes in orthorhombic Sm$_{5}$Ge$_{4}$-type structure.
The mean valence of Tm ions in Tm$_{5}$Si$_{2.0\pm 0.1}$Ge$_{2.0\pm 0.1}$ is almost trivalent. The 4f states is split by the crystalline electric field. The ground state exhibits the long range antiferromagnetic order with the ferromagnetically coupled magnetic moments in the $ac$ plane below  8.01 K,
while the exited states exhibit the reduction of magnetic moment and magnetic entropy and -log $T$-behaviors observed in Kondo materials.
\end{abstract}
 \maketitle

\section{Introduction}
Interest in R$_{5}$(Si$_{x}$Ge$_{1-x}$)$_{4}$ pseudobinary
compounds (R=rare-earth metals) has been revived with the recent
discovery by Pecharsky and
Gschneidner~\cite{Pecharsky971, Pecharsky972}of a giant magnetocaloric effect (MCE) in
Gd$_{5}$(Si$_{x}$Ge$_{1-x}$)$_{4}$. This effect has potential applications in magnetic refrigerants.
The giant MCE in Gd$_{5}$(Si$_{x}$Ge$_{1-x}$)$_{4}$ is caused by the
first-order magnetic transition that accompanies a martensitic-like
structure phase transition~\cite{Morellon98, Choe00}. In
Gd$_{5}$(Si$_{x}$Ge$_{1-x}$)$_{4}$, strong coupling between the
magnetic and crystallographic lattices is also responsible for
the giant magnetoresistance and strong magnetoelasic
effect~\cite{Morellon98, Levin99}. In order to understand the physical
mechanism for the interesting behaviors observed in the  R$_{5}$(Si$_{x}$Ge$_{1-x}$)$_{4}$ family of intermetallic compounds, it is
important to examine their magnetic and crystallographic structures.
The magnetic structure of Gd$_{5}$Si$_{2}$Ge$_{2}$ was determined by x-ray resonant magnetic scattering experiments~\cite
{Tan05}, However, the magnetic structure of
Gd$_{5}$(Si$_{x}$Ge$_{1-x}$)$_{4}$ has not been determined because of the huge neutron absorption cross section of the Gd
isotope. Therefore, the magnetic and thermal properties of many
binary and pseudobinary compounds of the form
R$_{5}$(Si$_{x}$Ge$_{1-x}$)$_{4}$, with rare-earth (R) ions excepting Gd ion are being
re-examined~\cite {Cadogan02, Rao04, Ritter02, Morellon03, Magen04, Cadogan04, Garlea05, Ritter06}.

Tb$_{5}$(Si$_{x}$Ge$_{1-x}$)$_{4}$ compounds also exhibit a
giant magnetocaloric effect around $x$=0.5~\cite {Morellon01}.
Initially, this phenomenon was understood to be the result of strong coupling of
the magnetic and crystallographic sublattices as in
Gd$_{5}$(Si$_{x}$Ge$_{1-x}$)$_{4}$ near $x$=0.5~\cite {Ritter02}.
However, no clear metamagnetic-like behavior, such as that found in
Gd$_{5}$Si$_{2}$Ge$_{2}$ was observed in the magnetization of
polycrystalline Tb$_{5}$Si$_{2}$Ge$_{2}$. A neutron powder
diffraction study of this compound revealed decoupling of the
structural and magnetic transitions with a separation of ~10 K~\cite
{Morellon03}. It was also revealed that, upon cooling, long-range ferromagnetism exists in the monoclinic $P112$$_1$$/a$
structure before its structural transformation into orthorhombic
$Pnma$.

On the other hand, Yb$_{5}$(Si$_{x}$Ge$_{1-x}$)$_{4}$ alloys
preserve the same crystal structure as $x$ varies from 0 to 4~\cite
{Ahn05}. Therefore, the replacement of Ge with Si and vice versa
has little effect on the magnetic properties of these materials,
which were thought to be a unique feature compared to the other
R$_{5}$T$_{4}$ systems. Three different lattice sites
accommodating lanthanides in the Gd$_{5}$Si$_{4}$-type crystal
structure exhibit selectivity with respect to the valence states of
Yb ions. Nonmagnetic Yb$^{2+}$ ions are located in the 4$c$
sites and one of the 8$d$ sites, whereas Yb$^{3+}$ ions are
located exclusively in the 8$d$ sites~\cite {Ahn05}.
Yb$_{5}$(Si$_{x}$Ge$_{1-x}$)$_{4}$ may be a
heterogeneous mixed valence system. All
Yb$_{5}$(Si$_{x}$Ge$_{1-x}$)$_{4}$ alloys show weak
antiferromagnetic correlations at temperatures between 2.4 K and 3.2
K~\cite {Ahn05}.

R$_{5}$(Si$_{x}$Ge$_{1-x}$)$_{4}$ systems with heavy lanthanides
other than Gd, Tb, and Yb (mentioned above) have been studies
to some extent. Recently, phase diagrams of the pseudobinary systems
with R=Er~\cite {Pecharsky04, Cadogan04} and Y~\cite {Pecharsky041} have been constructed. Selected
R$_{5}$(Si$_{x}$Ge$_{1-x}$)$_{4}$ compounds for R=La~\cite {Yang02, Yang03}, Ce~\cite {Gschneidner00}, Pr~\cite {Yang021, Yang031, Rao04}, Nd~\cite {Cadogan02, Yang022, Magen04}, Sm~\cite {Smith67}, Dy~\cite {Morellon04}, and Lu~\cite {
Ivtchenko00} have also been reported about their crystalline structures, magnetisms and magnetic caloric effects. Especially, although the Kondo effect is expected to be observed in the compounds with R=Ce~\cite {Gschneidner00} Yb~\cite {Ahn05} and Sm~\cite {Smith67}, there are no reports on the Kondo effect.

In case of R=Tm only the crystallographic data for the Tm$_{5}$T$_{4}$ binary
compounds (T=Si or Ge) have been reported~\cite {Smith671}, but their physical
properties have not been examined. In addition, since the Tm
can have valences of +2 or +3, the magnetic ordering, the valence fluctuation or the Kondo effect might be observed such as TmX (X: chalcogen) and thus this study on Tm$_{5}$Si$_{2}$Ge$_{2}$ is very important.

\begin{figure}[t] \centering
\includegraphics[width=1 \linewidth]{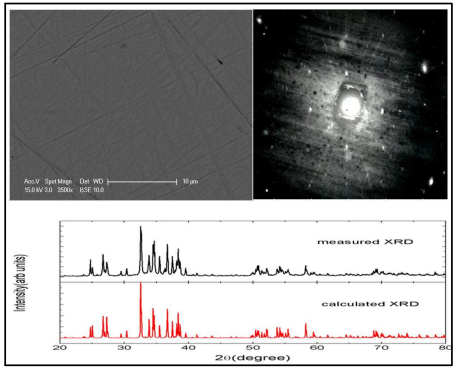} \caption{(color online).
(a) Scanning electron microscopy (SEM) image of the polished
surface of Tm$_{5}$Si$_{2}$Ge$_{2}$ single crystal. (b) Laue
diffraction pattern image in the ac-plane of a
Tm$_{5}$Si$_{2}$Ge$_{2}$ single crystal. (c) The measured and
calculated x-ray diffraction patterns of a Tm$_{5}$Si$_{2}$Ge$_{2}$
single crystal.} \label{figure1}\end{figure}

\section{Experimental Details}
A single crystal of Tm$_{5}$Si$_{2}$Ge$_{2}$ was grown using the
Bridgeman method at 1950 $^\circ$C in a pure tungsten crucible that had been previously baked out at 2200 $^\circ$C. The tungsten crucible never reacts with the constitutive elements. Because of the high volatility of Tm, the
starting materials were sealed in welded tungsten crucible using an
electron beam welder. Some Tm evaporates and might settle
somewhere else in the crucible. Therefore Tm, Si, and Ge at a ratio of 5.02 : 2 : 2 were
used as starting materials. The crystal had a diameter of 10 mm and a length of 12 mm, and
was well cleaved along the ac crystallographic plane. Tm metal was
obtained from a commercial vendor and was 99.9 at.$\%$ pure with
the following major impurities (in at.$\%$): Fe-0.01, Ca-0.03, Mg-0.01,
Ni-0.01, Al-0.01, T-0.01, Si-0.01, C-0.015, O-0.05,and Cl-0.05. The
elements Si and Ge were also purchased from a commercial vendor, and
were better than 99.999 at.$\%$ purity. The orientation of the $b$
crystallographic axis and $ac$ plane in the sample were
established using the backscattered Laue method.

The magnetic measurements were performed using a SQUID magnetometer
(MPMS XL, Quantum Design). The magnetic susceptibility of
the zero-magnetic-field cooled and magnetic-field cooled samples was
measured as a function of temperature from 2 to 300 K at $H$=100 Oe.
Isothermal magnetization was measured at 2, 4, and 6 K in a DC
magnetic field varying from 0 to 5 T for the zero-magnetic-field
cooled and magnetic-field cooled samples. The specific heats of
Tm$_{5}$Si$_{2}$Ge$_{2}$ and Lu$_{5}$Si$_{2}$Ge$_{2}$ were measured
using a physical property measurement system (PPMS, Quantum Design).
The former was measured in various magnetic fields, ranging from 0 to 9
T, and oriented parallel to the $b$ crystallographic axis.

\begin{figure}[t] \centering
\includegraphics[width=1 \linewidth]{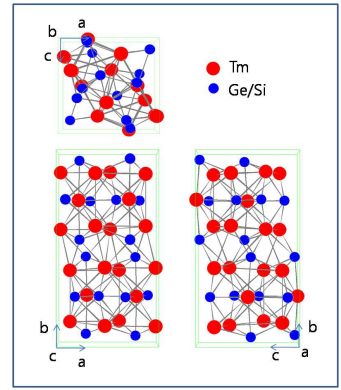} \caption{(color online).
Orthorhombic Sm$_{5}$Ge$_{4}$-type crystal structure with space group $Pnma$} \label{figure2}\end{figure}

The mean valence of Tm-ions in Tm$_5$Si$_2$Ge$_2$ was examined by X-ray photoemission spectroscopy (XPS, Mg $K\alpha$ line,
$h\nu$=1253.6~eV) of the Tm $4d$ and
$4p$ core levels using a 100-mm radius hemispherical photoelectron analyzer
(VG Scienta SES-100).
The base pressure of the chamber was less than $2\times10^{-8}$~Pa.
The sample temperature and total energy resolution were set to approximately 20~K and 0.7~eV, respectively.
To obtain the mean valence of Tm$_5$Si$_2$Ge$_2$, the Tm $4d$ and $4p$ core
level XPS spectra of TmS (mostly Tm$^{3+}$) and TmTe (mostly Tm$^{2+}$)
were also measured for the reference.~\cite{Nath2003}
Clean sample surfaces were prepared inside the ultra high vacuum chamber
by scraping with a clean diamond filler.
After cleaning, the level of oxygen and carbon contaminations was checked by
monitoring the intensity of the O~$1s$ and C~$1s$ photoemission peaks.
The intensities of the O~$1s$ and C~$1s$ peaks were kept within the noise level
during the measurements.

\begin{figure}[t] \centering
\includegraphics[width=1 \linewidth]{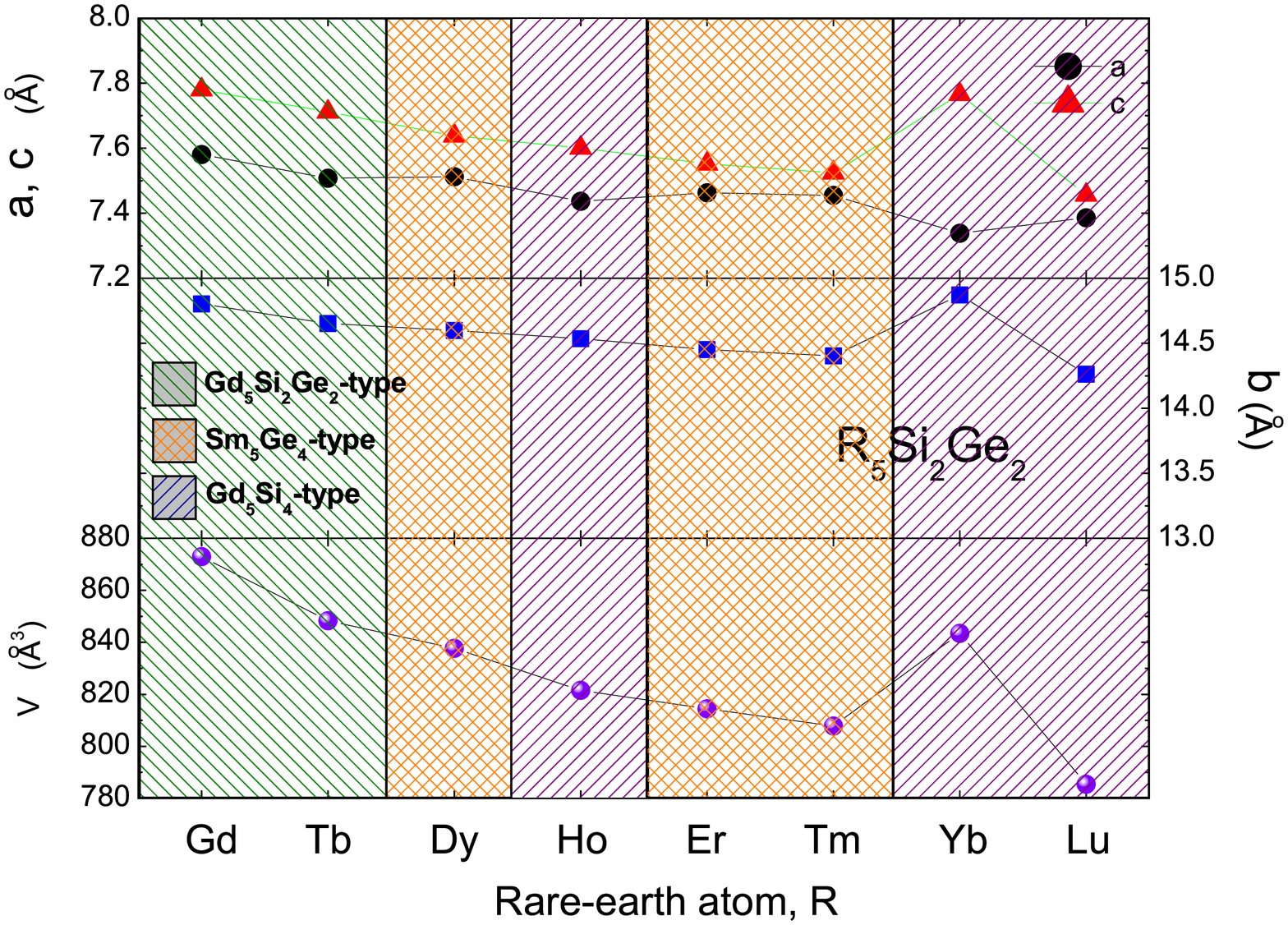} \caption{(color online).
Variation of the lattice parameters of R$_{5}$Si$_{2}$Ge$_{2}$ with
R=Gd through to Lu} \label{figure3}\end{figure}

\section{Experimental Results and Discussion}
As shown in Fig. 1(a), atomic number sensitive back-scattered
electron (BSE) imaging of the polished surface of single crystal
Tm$_{5}$Si$_{2}$Ge$_{2}$ showed that the sample contained a single
phase, which was identified as Tm$_{5}$Si$_{2.0\pm 0.1}$Ge$_{2.0\pm 0.1}$ by EDS (Energy Dispersive Spectroscopy)
. Fig. 1(b) shows the imaging of the pattern of Laue
diffraction for single crystal Tm$_{5}$Si$_{2}$Ge$_{2}$. The Laue pattern was formed the x-rays
diffracted from the ac-plane of Tm$_{5}$Si$_{2}$Ge$_{2}$. Fig. 1(c)
shows the x-ray diffraction pattern of Tm$_{5}$Si$_{2}$Ge$_{2}$. The
peaks positions in the pattern are consistent with the diffraction
pattern calculated from an orthorhombic Sm$_{5}$Ge$_{4}$-type
crystal structure (space group $Pnma$), which is plotted in Fig. 2, with lattice constants
$a$=7.455$\pm$0.001, $b$=14.402$\pm$0.002, and $c$=7.525$\pm$0.001 {\AA}.

\begin{figure}[t] \centering
\includegraphics[width=1 \linewidth]{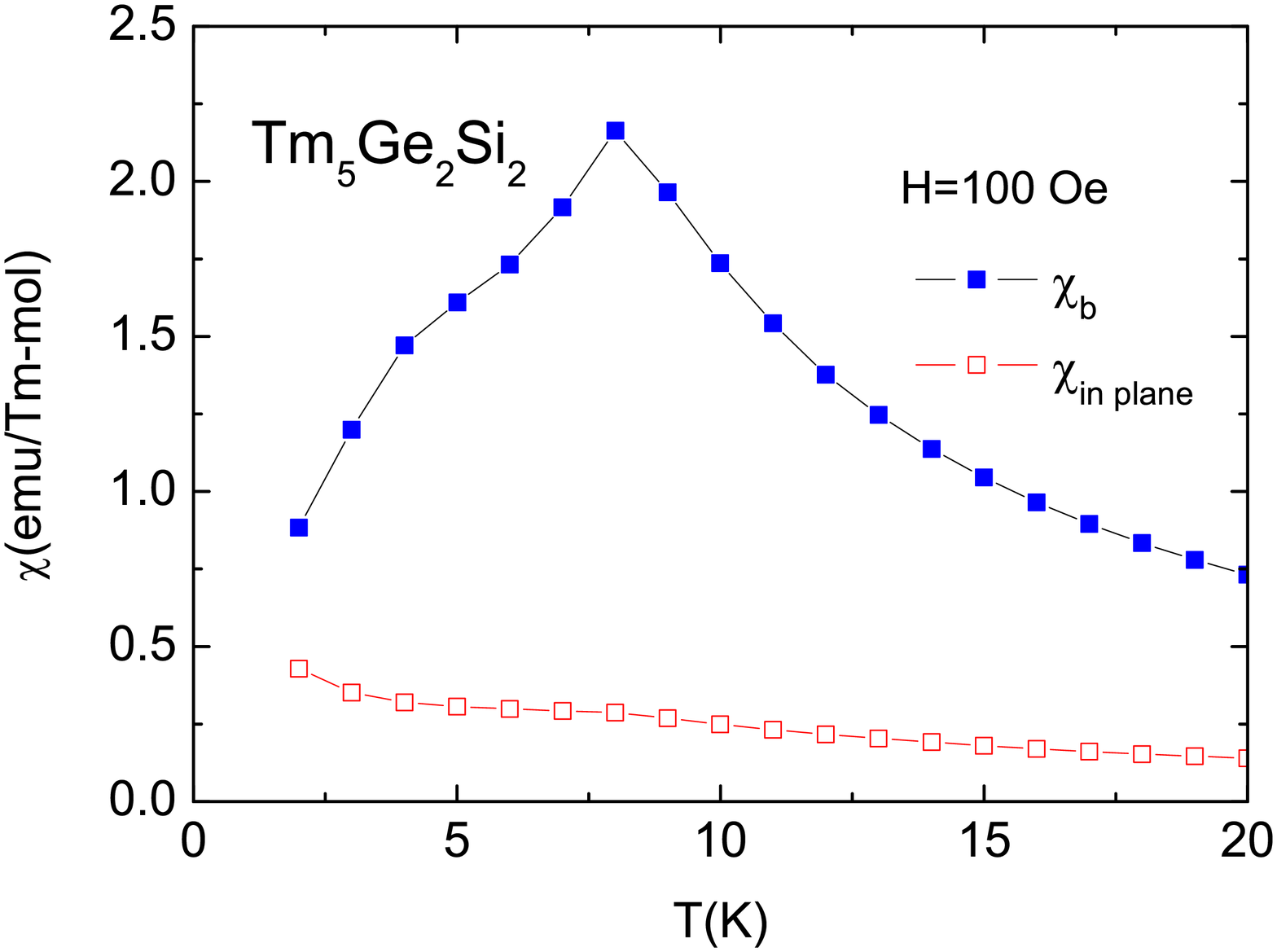} \caption{(color online).
Temperature dependence of the magnetic susceptibility of a
Tm$_{5}$Si$_{2}$Ge$_{2}$ single crystal for the
${H}${$\parallel$}$b$-axis and {$H$}{$\parallel$}$ac$-plane in low
temperature regions.}
\label{figure4}\end{figure}

Fig. 3 shows the change in the
lattice parameters of R$_{5}$Si$_{2}$Ge$_{2}$ with R=Gd through to Lu.
R$_{5}$Si$_{2}$Ge$_{2}$ is classified into three types of
crystal structures, which makes it difficult to deduce the mixed valence of
Tm$_{5}$Si$_{2}$Ge$_{2}$. It was possible to estimate the valence of R-ions because the three types of crystal structures are quite
similar and contain four formula units in an unit cell. In R=Gd to Er and Lu,
which are trivalent, the lattice parameters and volume of a
R$_{5}$Si$_{2}$Ge$_{2}$ unit cell decrease smoothly with increasing atomic number
due to lanthanide contraction. In the case of R=Tm, the lattice parameters
and unit cell volume were on the line indicating that Tm ion is trivalent.
This result is inconsistent with those of the magnetic susceptibility and
magnetization mentioned below. Note that the unit cell volume for R=Yb is
smaller than that indicated from the line formed by the trivalent R ion,
which means that Yb ions have an intermediate valence, as reported elsewhere.

\begin{figure}[t] \centering
\includegraphics[width=1 \linewidth]{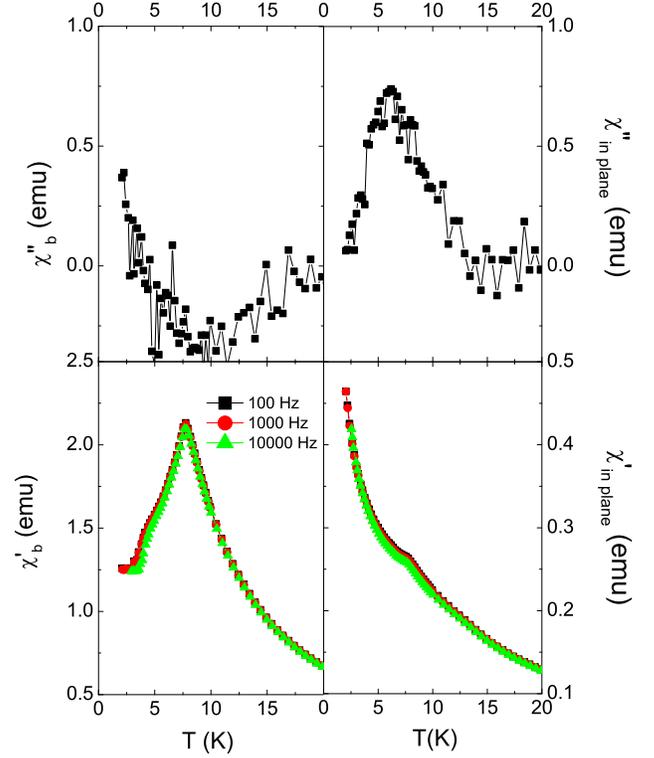} \caption{(color online).
Ac magnetic susceptibilities as a function of temperature for
different frequencies in Tm$_{5}$Si$_{2}$Ge$_{2}$ single
crystal.} \label{figure5}
\end{figure}

Fig. 4 shows the temperature dependence of the magnetic susceptibility $\chi$ of a single crystal of
Tm$_{5}$Si$_{2}$Ge$_{2}$ in the low temperature region. The magnetic susceptibilities were measured
upon heating at $H$=100 Oe with a magnetic field aligned
parallel to the $b$ crystallographic axis and the $ac$ plane, i.e.
$\chi$$_{b}$ and $\chi$$_{in-plane}$, respectively. The magnetic susceptibilities of
the zero-magnetic-field cooled and field cooled samples were the same.
$\chi$$_{b}$ shows a distinct peak at 8.0~K, which is due to antiferromagnetic ordering. Below $T_N$, $\chi$$_{b}$ approached
zero, and then showed a shoulder at approximately 5~K. On the other hand,
the $\chi$$_{in-plane}$ below $T_N$ was a constant down to approximately 5~K,
and then increased with decreasing temperature. The
behaviors of the magnetic susceptibility approaching zero in $\chi$$_{b}$ and being constant in
the $\chi$$_{in-plane}$ immediately below $T_N$ are also found in the conventional
antiferromagnetic single crystals. This indicates that the magnetic
moments of Tm$_{5}$Si$_{2}$Ge$_{2}$ below $T_N$ are coupled
atiferromagnetically along the $b$-axis. The shoulder in
the $\chi$$_{b}$ and the increase in the $\chi$$_{in-plane}$ at approximately 5~K were
due to the magnetic moment coupled ferromagnetically in the plane,
which will be discussed in the section reporting the ac magnetic susceptibility. Such
antiferromagnetism was proposed by Landau for layered
antiferromagnetisc materials in which the magnetic moments in
ferromagnetically-ordered layers alternate from layer to layer~\cite
{Held97}. In Gd$_{5}$Ge$_{4}$ the antiferromagnetism proposed by
Landau was also observed at $T$$_{N}$=128 K, and the strength of the
exchange interactions for antiferromagnetism and ferromagnetism were
equal~\cite {Levin04}. However, there was anisotropy in the strength of the exchange
interactions in Tm$_{5}$Si$_{2}$Ge$_{2}$.

\begin{figure}[t] \centering
\includegraphics[width=1 \linewidth]{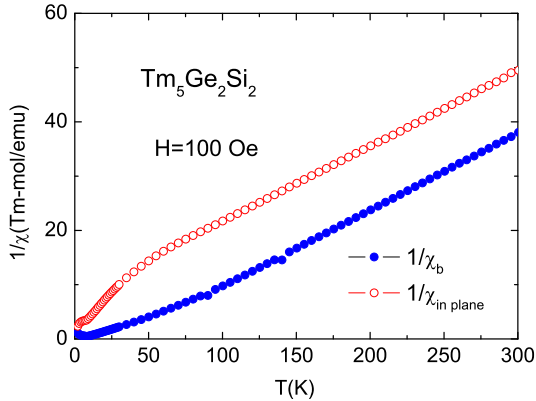} \caption{(color online).
Temperature dependence of the inverse magnetic susceptibility of a
Tm$_{5}$Si$_{2}$Ge$_{2}$ single crystal for the
${H}${$\parallel$}$b$-axis and {$H$}{$\parallel$}$ac$-plane.}
\label{figure6}\end{figure}

To prove our claim regarding the layered magnetic orderings
obseved in Tm$_{5}$Si$_{2}$Ge$_{2}$ the ac magnetic
susceptibilities were measured as a function of temperature at different
frequencies (Fig. 5). In $\chi$$_{b}$$'$, a peak was observed at 8~K and
a bulge was observed at approximately 5~K, which is similar to the dc magnetic
susceptibility. No frequency-dependence was found in the anomalies
formed by the magnetic ordering, which suggests that the anomalies are
caused by long range magnetic orderings. $\chi$$_{b}$$''$ is
independent of temperature below $T_N$, which suggests that magnetic ordering is not ferromagnetic. On the other hand,
the $\chi$$_{in-plane}$$'$ below $T_N$ is constant down to approximately 5~K, and then increases with decreasing temperature, which is similar to the
dc magnetic susceptibility. The $\chi$$_{in-plane}$$''$ shows a peak
at approximately 5~K, which is different from the feature in $\chi$$_{b}$$'$. The
peak originates from a long range ferromagnetic ordering in the plane below
5~K because the peak in the imaginary part of the magnetic
susceptibility is formed by energy losses due to the hysteresis observed
in ferromagnetic ordering. This strongly supports the claim for
the magnetic ordering mentioned above.

Fig. 6 shows the temperature dependence of the inverse magnetic
susceptibility 1/$\chi$. 1/$\chi$ is linearly proportional to the temperature at temperature regions
higher than 150K, and reflects the Curie-Weiss behavior with the
paramagnetic Curie temperatures ${\Theta}$$_{p,b}$$\sim$34 K and
$\Theta$$_{p, in-plane}$$\sim$-55 K in the $b$-axis and the plane,
respectively. The effective magnetic moment, $p$$_{eff}$, is equal
to 6.8$\pm$ 0.15 $\mu$$_{B}$/Tm atom in both directions. The anisotropy of
$\Theta$$_{p}$ may be due to the anisotropy of the exchange
interactions or the crystalline electrical effect. The
experimentally determined $p$$_{eff}$ was between that of Tm$^{3+}$({$p$}$_{eff
1}$=7.57 $\mu$$_{B}$) and Tm$^{2+}$($p$$_{eff 2}$=4.54 $\mu$$_{B}$), which were determined theoretically calculated.

\begin{figure}[t] \centering
\includegraphics[width=1 \linewidth]{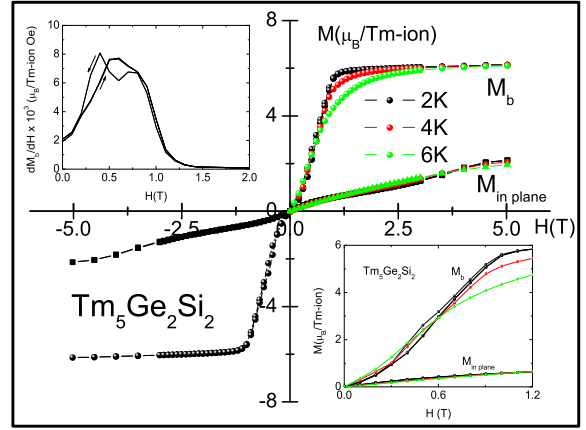} \caption{(color online).
Magnetic field dependence of the magnetization of a
Tm$_{5}$Si$_{2}$Ge$_{2}$ single crystal for the
$H$$\parallel$$b$-axis and $H$$\parallel$$ac$-plane, $M$$_{b}$ and
$M$$_{in-plane}$, respectively, at $T$= 2, 4, and 6 K. The inset
plotted below shows $M$$_{b}$ and $M$$_{in plane}$ in low magnetic
fields, and the inset plotted above shows the derivatives of
$M$$_{b}$ with respect to the magnetic field at $T$=2~K.}
\label{figure7}
\end{figure}

The isothermal magnetization of single crystal
Tm$_{5}$Si$_{2}$Ge$_{2}$ were measured at T=2, 4, and 6 K, as shown in Fig. 7, when a magnetic field was applied in the directions
parallel to the $b$ crystallographic axis and $ac$ plane. The
magnetization of the $H$$\parallel$$b$ axis, $M$$_{b}$, was much
larger than that of the $H$$\parallel$$ac$ plane, $M$$_{in-plane}$,
in the measured magnetic field regions, which suggests that the $b$
axis is the easy axis. A metamagnetic transition below $T$$_{N}$ was observed at
$H$$_{cr}$=0.5 T, as shown in the inset plotted in the bottom
of Fig. 7. This is better seen as the peak in another inset,
which shows the derivatives of the magnetization with respect to
the magnetic field. The metamagnetic transition was attributed to a magnetic
field induced spin-flip transformation. Remanence-free
hysteresis was observed only at $T$=2 K. As shown in the bottom
inset, $M$$_{b}$ increases with increasing temperature below
$H$$_{cr}$. Hysteresis is generally observed in magnetic
materials with narrow domain walls~\cite {Niraj05, Niraj07, Wang06, Nirmala07}. In these magnetic materials, the motion of the
walls is hindered by pinning centers at low temperatures, giving
rise to small magnetization. As the temperature increases, thermal
energy provides the driving force necessary to overcome the barriers
created by the pinning centers, leading to an increase in magnetization with increasing temperature.
The $M$$_{b}$ remains below 6.1$\pm$ 0.1 $\mu$$_{B}$ per Tm-ion in a
magnetic field of 5 T. The theoretically saturated magnetic moments of
Tm$^{3+}$ and Tm$^{2+}$ are given by $gJ$=7 $\mu$$_{B}$, $M$$_{3+}$,
and $gJ$=4 $\mu$$_{B}$, $M$$_{2+}$, respectively, where $g$ is the
gyromagnetic ratio and $J$ is the total angular momentum quantum
number. An intermediate magnetization value was also observed in the magnetically ordered
state. $M$$_{in-plane}$ in the
$ac$ plane increases slowly with increasing magnetic field with no distinct change at $H$$_{cr}$. This indicates that the
metamagnetic transition is due to a flip of the magnetic moments in
the ac plane.

\begin{figure}[t] \centering
\includegraphics[width=1 \linewidth]{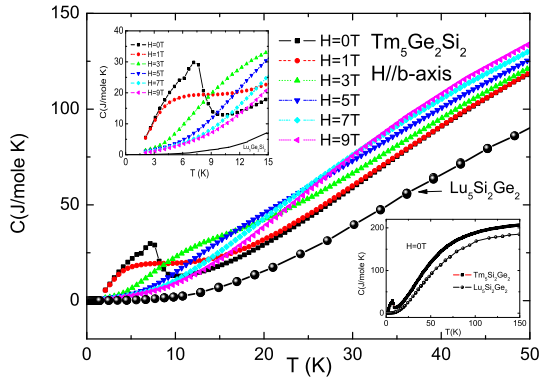} \caption{(color online).
Temperature dependence of the specific heats, $C$, of a
Tm$_{5}$Si$_{2}$Ge$_{2}$ single crystal in various magnetic fields
for the $H$$\parallel$$b$-axis and Lu$_{5}$Si$_{2}$Ge$_{2}$. The
left inset shows the specific heat in low temperature regions and the right inset shows the specific heat in measured temperature regions.}
\label{figure8}
\end{figure}

\begin{figure}[t] \centering
\includegraphics[width=1 \linewidth]{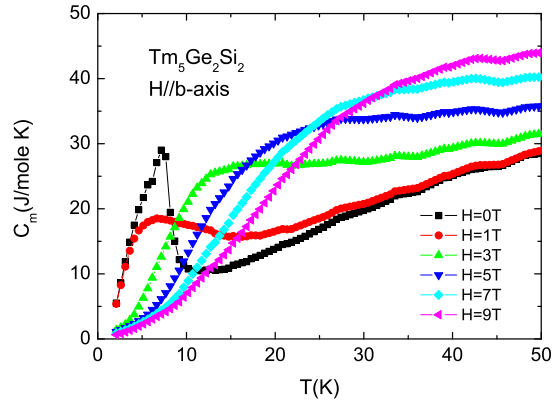} \caption{(color online).
Temperature dependence of the magnetic contribution ($C$$_{m}$) to
specific heats of Tm$_{5}$Si$_{2}$Ge$_{2}$ single
crystal in various magnetic fields for the $H$$\parallel$$b$-axis.
}\label{figure9}
\end{figure}

The specific heat of the single crystal of Tm$_{5}$Si$_{2}$Ge$_{2}$
was measured upon heating in various magnetic fields ranging from 0 to
9 T, which is shown in Fig. 8. The magnetic field was applied
parallel to the $b$ crystallographic axis. The antiferromagnetic
transition which was observed at $T$$_{N}$=8.0~K in the magnetic
susceptibility measurements taken at $H$=100 Oe is indicated by a
sharp peak at 7.7~K in the specific heat measurements at $H$=0 T.
The mid-point temperature (8~K) of the peak in the specific heat
curve is equal to $T$$_{N}$. When $H$$>$1 T, the sharp peak becomes
a broad shoulder and shifts to high temperature with further
increasing in $H$. To demonstrate this feature distinctly, the phonon part in
the specific heat of Tm$_{5}$Si$_{2}$Ge$_{2}$ was excluded using the
specific heat of the nonmagnetic Lu$_{5}$Si$_{2}$Ge$_{2}$. Fig. 9 shows the magnetic contribution to the specific heat. In the figure, there appears to several small humps at approximately $T$=26, 33, 42~K etc.
These humps are due to errors occurring when subtracting the specific heat of
Lu$_{5}$Si$_{2}$Ge$_{2}$, which were measured at rough temperature intervals.
A shoulder was observed at approximately 7 K at $H$=1 T,
12 K at $H$=3 T, 18 K at $H$=5 T, 28 K at $H$=7 T, and 35 K at $H$=9
T. The shoulder was formed by a Schottky anomaly due to excitation
between the crystal-field splitting states, which are divided by the
Zeeman effect under internal magnetic fields induced in
ferromagnetism as well as in an applied magnetic field~\cite{Blanco91, Bouvier91, Sampathkumaran98}.
A weak anomaly was observed at 5.6 K
in the specific heat curve at $H$=0 T, as shown in the inset of Fig. 8. This was attributed to the ferromagnetic ordering in the plane observed in the magnetic
susceptibility. This anomaly was not obseved in $H$$>$1 T because
Tm$_{5}$Si$_{2}$Ge$_{2}$ exhibits ferromagnetism.
Fig. 10 shows the magnetic entropy per Tm mole evaluated from the integral of
the magnetic contribution to the specific heat divided by temperature. The magnetic entropy at $H$$=$0 T was nearly recovered to $R$ln2 at $T_N$. This suggests that the ground state is a doublet considering the low symmetry of its crystal structure.
In the high temperature regions, the magnetic entropy was
saturated to approximately 17.5 J/Tm-molK$^2$, which is smaller than the full
entropy for Tm$^{3+}$ ions, $R$ln13. This is due to the reduced magnetic moment observed
in the magnetic susceptibility and magnetization. The magnetic entropy in $H>$0 T
approaches the value at $H$$=$0 T in the high temperature regions.
The entropy decreases with increasing magnetic field across all temperatures,
because of the Zeeman effect mentioned above.

\begin{figure}[t] \centering
\includegraphics[width=1 \linewidth]{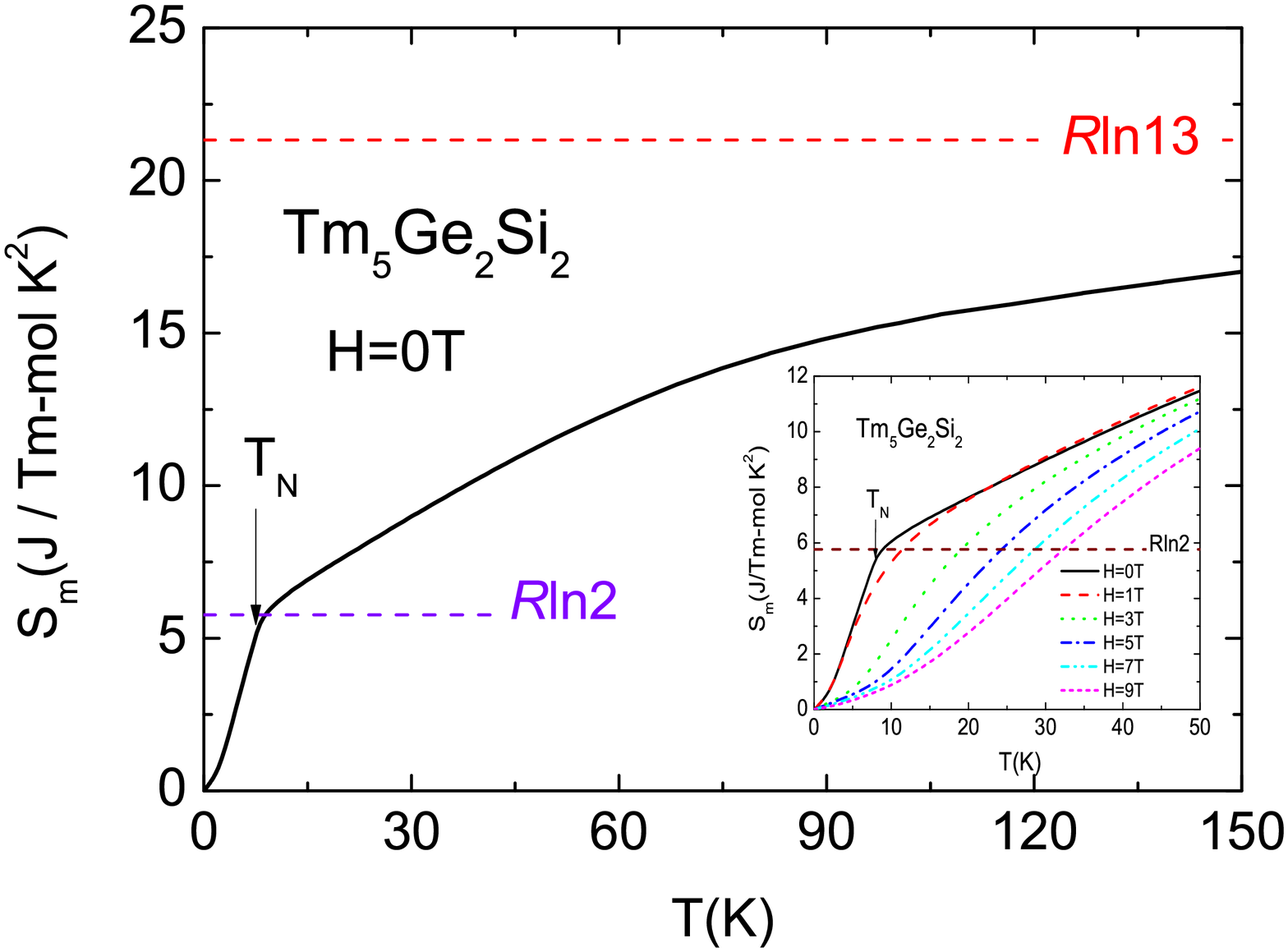} \caption{(color online).
Temperature dependence of the magnetic contribution ($S$$_{m}$) to
the entropy of Tm$_{5}$Si$_{2}$Ge$_{2}$ single crystal at $H$=0 T. $R$ln13
and $R$ln2 are the full entropies of Tm$^{+3}$ and the entropy of a doublet ground state, respectively. The insect shows the entropy under magnetic fields.}\label{figure10}
\end{figure}

However, it is unclear why the magnetic moment is reduced. Decreases in the magnetic moments of Tm ions might arise from the mixed valent state, spin
frustration, Kondo effect etc. Assuming that there are two distinct Tm valence states in the
lattice, the fraction of each ion can be estimated from the experimentally evaluated effective magnetic moment using the
following equation: $p$$_{eff}$=[$\alpha$$\cdot$$p$$_{eff 1}$$^{2}$ +
(1-$\alpha$)$\cdot$$p$$_{eff 2}$$^{2}$]$^{1/2}$, where $\alpha$ is the
fraction of Tm$^{3+}$ ions. Solving with respect to $\alpha$, the
fraction of Tm$^{3+}$ and Tm$^{2+}$ ions in the unit cell is 0.70$\pm$ 0.05 and 0.30$\pm$ 0.05, respectively. The fraction of each ion present can be estimated from the magnetization in the magnetically ordered state using the
following equation: $M$=[$\alpha$$\cdot$$M$$_{3+}$ +
(1-$\alpha$)$\cdot$$M$$_{2+}$]. Solving with respect to $\alpha$, the fraction of Tm$^{3+}$ and Tm$^{2+}$ ions in
the unit cell is 0.70$\pm$ 0.03 and 0.30$\pm$ 0.03, respectively. The value
of $\alpha$ obtained from the saturating magnetic moment is equal to that obtained from the above mentioned effective magnetic moment. However, the mixed valent state differs from the result of the change in lattice constants.

\begin{figure}[t]
\begin{center}
\includegraphics[width=1\linewidth]{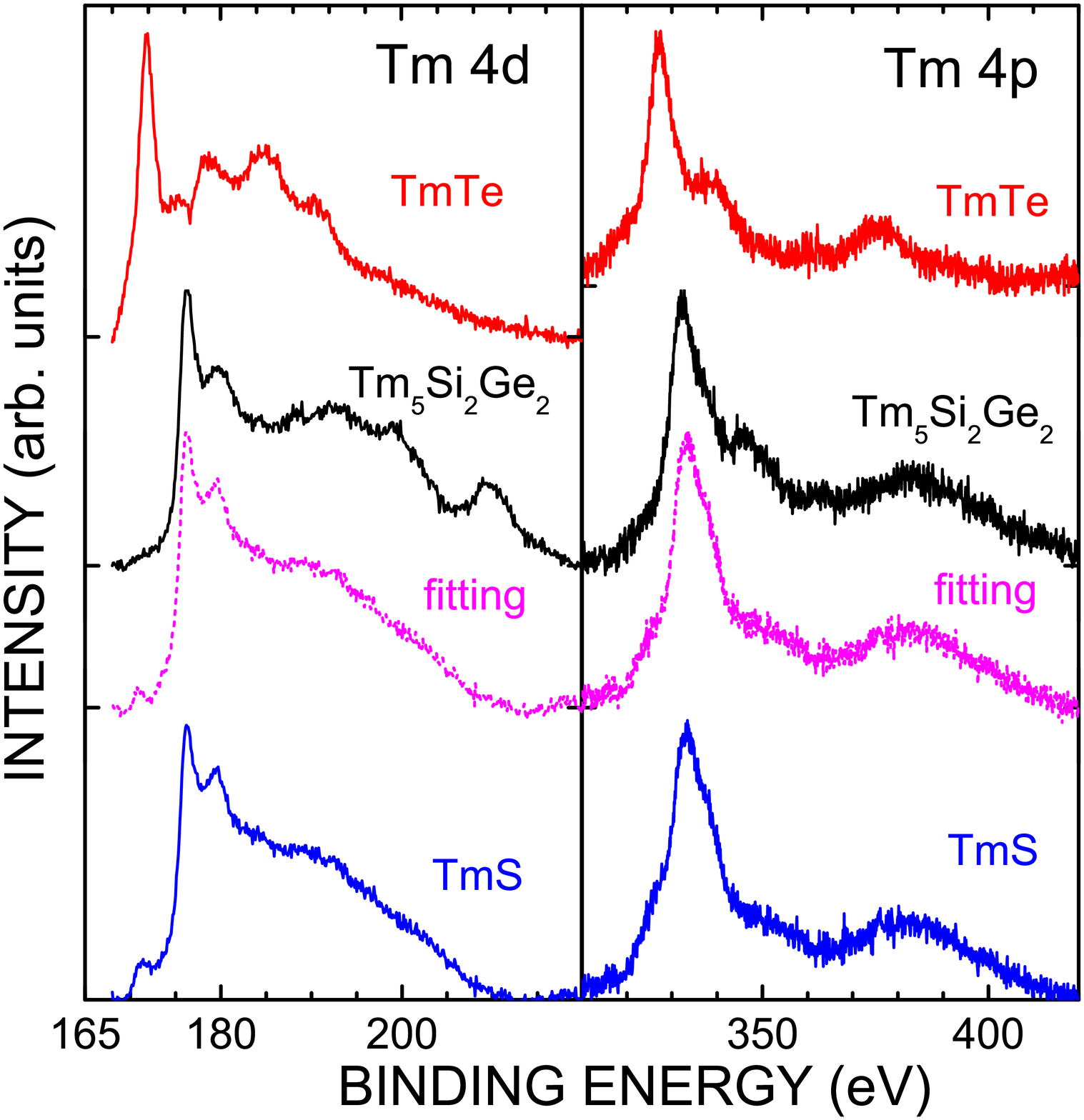}
\end{center}
\caption{
(Color online)
Tm~$4d$ and $4p$ XPS spectra of Tm$_5$Si$_2$Ge$_2$ and the reference materials, TmS and TmTe measured at $T$~=~20~K.
The fitting results using the linear combination of $1.1\cdot I_{\rm TmS}(E)-0.1\cdot I_{\rm TmTe}(E)$ are also plotted.}\label{figure11}
\end{figure}

In order to confirm the mixed valent state in more detail, the Tm~$4d$ and $4p$
XPS spectra of Tm$_5$Si$_2$Ge$_2$ was measured, as shown in Fig. 11.
The XPS spectra of TmS and TmTe at the same core levels are also plotted as
a reference for the Tm$^{3+}$ and Tm$^{2+}$ spectra, respectively.
The Tm~$4d$ and $4p$ XPS spectra of TmS have shoulders at the binding energies of 172
and 325~eV, respectively, which originate from the Tm$^{2+}$ state appearing in the XPS spectra of TmTe.
These spectral features are consistent with the mixed valent nature of TmS.~\cite{Bucher1975}
In Tm$_5$Si$_2$Ge$_2$, since the Tm$^{2+} 4d$ peaks do not appear in the both Tm~$4d$
and $4p$ XPS spectra, the mean valence is almost trivalent compared to TmS.
The Tm~$4d$ and $4p$ XPS spectra of Tm$_5$Si$_2$Ge$_2$ can be fitted by a linear
combination of the XPS spectra of TmS and TmTe using the following function:
$I_{\rm Tm_5Si_2Ge_2}(E)=\alpha \cdot I_{\rm TmS}(E)+(1-\alpha)\cdot I_{\rm TmTe}(E)$.
Here $I_{sample}(E)$ indicates the XPS spectrum and $\alpha$, which is the fraction of
TmS, becomes $1.1\pm0.1$.
Since the mean valences of Tm ions in TmS and TmTe are 2.8 and 2.0, respectively,~\cite{Nath2003}
the mean valence of Tm$_5$Si$_2$Ge$_2$, $z$, can be estimated to be approximately $2.9\pm0.1$
using the function $z=0.8\cdot\alpha+2.0$.
Therefore, the mean valence of Tm ions in Tm$_5$Si$_2$Ge$_2$ evaluated from the XPS
spectra was almost trivalent, which is consistent with the change in unit cell volume. The change in the lattice constants and
XPS are more powerful than the magnetic susceptibility and magnetization because the
formers are the results observed directly from a mixed valence. In this context, the reduced magnetic moment of Tm ions in Tm$_{5}$Si$_{2}$Ge$_{2}$ is not due to the mixed valent state.

Spin frustration is often observed in triangular,
Kagome and pyrochlore crystal structures. Tm$_{5}$Si$_{2}$Ge$_{2}$ does not contain
these structures, which excludes the spin frustration.

On the other hand, the phenomenon, in which conduction electrons strongly couple with f-electron spin in the opposite direction
by a c-f interaction is known as the Kondo effect. The Kondo effect often causes a decrease in magnetic moment and the magnetic entropy.
To confirm the correlation between these phenomena and the Kondo effect, this study measured
the electrical resistivity of Tm$_{5}$Si$_{2}$Ge$_{2}$ and Lu$_{5}$Si$_{2}$Ge$_{2}$, which does not contain 4f electrons and is a reference system for examining magnetic transport in Tm$_{5}$Si$_{2}$Ge$_{2}$. Fig. 12 shows their electrical resistivities. Lu$_{5}$Si$_{2}$Ge$_{2}$ showed a normal metallic temperature dependence, while Tm$_{5}$Si$_{2}$Ge$_{2}$ showed two anomalies below 10~K and approximately 200~K. To observe the anomalies in detail, the electrical resistivity of Lu$_{5}$Si$_{2}$Ge$_{2}$ was subtracted from that of Tm$_{5}$Si$_{2}$Ge$_{2}$. The result is shown in Fig. 13. The anomalies revealed a log $T$-dependence. The log $T$-behavior is a characteristic of the Kondo effect. The log $T$-behavior observed in the low temperature regions begins from near $T_N$. In addition, the magnetic
\begin{figure}[t] \centering
\includegraphics[width=1 \linewidth]{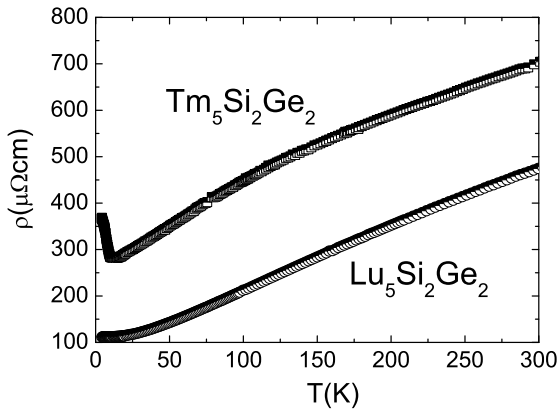} \caption{(color online).
Temperature dependence of the electrical resistivity of Tm$_{5}$Si$_{2}$Ge$_{2}$ single crystal and Lu$_{5}$Si$_{2}$Ge$_{2}$ poly crystal.}\label{figure12}
\end{figure}
entropy of Tm$_{5}$Si$_{2}$Ge$_{2}$ recovered to the full entropy of the ground doublet
at $T_N$. This indicates that the ground doublet state, which comes from the crystalline electric field (CEF) splitting for the Hund's rule ground state of $J$=6, is not affected by the Kondo effect, whereas the excited states near the ground state are affected by the Kondo effect. The log $T$-behavior in high temperature regions is due \begin{figure}[t] \centering
\includegraphics[width=1 \linewidth]{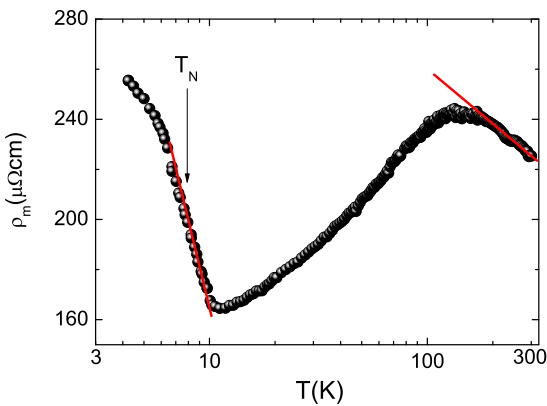} \caption{(color online).
Temperature dependence of the magnetic resistivity of Tm$_{5}$Si$_{2}$Ge$_{2}$ subtracting the resistivity of Lu$_{5}$Si$_{2}$Ge$_{2}$.}\label{figure13}
\end{figure}
to the Kondo effect for more excited CEF states. The Kondo temperature ($T_K$) was estimated to be higher than 200~K according to the Kondo model. The high $T_K$ interrupts the recovery of magnetic entropy at high temperatures, as mentioned above.
The magnetic moments of the excited states screened by the Kondo effect
should cause lower decrease in effective magnetic moment and magnetization. The Kondo effect is caused by the hybridization between conduction electrons and 4f electrons. the strength of hybridization depends on the overlap of their wavefunctions. When the excited crystal field orbits have a higher overlap with orbits of 5d conduction electron than the ground orbits, the Kondo effect due to the excited states dominates. This was well studied in Ce monopnictides~\cite{Takahashi85}.

\section{Conclusion}

Tm$_{5}$Si$_{2}$Ge$_{2}$ crystallizes in an orthorhombic Sm$_{5}$Ge$_{4}$-type
structure at 300 K. The long range antiferromagnetic order coupled along
the $b$ crystallographic axis is found below $T$$_{N}$=8.01
K. Below about 5 K the antiferromagnetic
magnetic moments coupled along the $b$ crystallographic axis remain
and the magnetic moments in the $ac$ crystallographic plane are
coupled ferromagnetically. The reduction of magnetic moment and magnetic entropy and -log $T$ dependence in electrical resistivity were observed in high temperature regions due to the Kondo effect on excited crystal field states. The computation between the Kondo effect and magnetic order plays an important role in Tm$_{5}$Si$_{2}$Ge$_{2}$.

\begin{acknowledgments}
 This work was performed for the Nuclear R$\&$D Programs funded
by the Ministry of Science $\&$ Technology (MOST) of Korea.
\end{acknowledgments}

\end{document}